\documentclass[aps,prb,superscriptaddress,twocolumn]{revtex4-1}
\pdfoutput=1
\usepackage{amsmath}
\usepackage{ amssymb }
\usepackage{graphicx}
\usepackage{color}

\begin{document}

\newcommand{\beq}{\begin{equation}}
\newcommand{\eeq}{\end{equation}}
\newcommand{\beqa}{\begin{eqnarray}}
\newcommand{\eeqa}{\end{eqnarray}}
\newcommand{\ben}{\begin{enumerate}}
\newcommand{\een}{\end{enumerate}}
\newcommand{\hs}{\hspace{0.5cm}}
\newcommand{\vs}{\vspace{0.5cm}}
\newcommand{\note}[1]{{\color{red} #1}}
\newcommand{\tim}{$\times$}
\newcommand{\bigo}{\mathcal{O}}
\newcommand{\bra}[1]{\ensuremath{\langle#1|}}
\newcommand{\ket}[1]{\ensuremath{|#1\rangle}}
\newcommand{\bracket}[2]{\ensuremath{\langle#1|#2\rangle}}

%\title{Universality in R\'enyi entanglement entropies at a strongly-interacting quantum critical point}
\title{Corner contribution to the entanglement entropy of strongly-interacting \\ O(2) quantum critical systems in 2+1 dimensions} 

\author{E.M.\ Stoudenmire}
\affiliation{Perimeter Institute for Theoretical Physics, Waterloo, Ontario, N2L 2Y5, Canada}

\author{Peter\ Gustainis}
\affiliation{Perimeter Institute for Theoretical Physics, Waterloo, Ontario, N2L 2Y5, Canada}
\affiliation{Department of Physics and Astronomy, University of Waterloo, Ontario, N2L 3G1, Canada}

\author{Ravi\ Johal}
\affiliation{Department of Physics and Astronomy, University of Waterloo, Ontario, N2L 3G1, Canada}

\author{Stefan\ Wessel}
\affiliation{Institut f\"ur Theoretische Festk\"orperphysik, JARA-FIT and JARA-HPC,RWTH Aachen University, 52056 Aachen, Germany}

\author{Roger G.\ Melko}
\affiliation{Perimeter Institute for Theoretical Physics, Waterloo, Ontario, N2L 2Y5, Canada}
\affiliation{Department of Physics and Astronomy, University of Waterloo, Ontario, N2L 3G1, Canada}

\date{\today}

\begin{abstract}
In a $D=2+1$ quantum critical system, the entanglement entropy across a boundary with a corner 
contains a subleading logarithmic scaling term with a universal coefficient. 
It has been conjectured that this coefficient is, to leading order, proportional 
to the number of field components $N$ in the associated $O(N)$ continuum $\phi^4$ field theory.
Using density matrix renormalization group calculations combined with the powerful numerical linked cluster 
expansion technique, we confirm this scenario for the $O(2)$ Wilson-Fisher fixed point 
in a striking way, through direct calculation at the quantum critical points of two very different microscopic models.
The value of this corner coefficient is, to within our numerical precision, twice the coefficient of the Ising fixed point.
Our results add to the growing body of evidence that this universal term in the R\'enyi entanglement entropy
reflects the number of low-energy degrees of freedom in a system, even for strongly interacting theories.
\end{abstract}

\maketitle

\section{Introduction}

In an information-theoretic sense, it is no surprise that quantities related to the entropy of a many-body system reflect the underlying
degrees of freedom in the system---a fact related even to the foundations of statistical thermodynamics.
In recent years, this intuition has revealed new connections between complex interacting quantum many-body systems at a quantum
critical point and the field theories capturing their universal low-energy behavior.
These connections are made through the quantum system's {\it entanglement} entropy, measured across a bipartition between
two subregions.  Since, for space-time dimensions $D$ higher than 1+1,
the entanglement entropy is dominated by a non-universal term proportional to the size of the 
bipartition boundary (the ``area'' law), one typically must search for such universal quantities in the coefficents of sub-leading
scaling terms, which themselves depend on the subregion geometry and topology.\cite{Fradkin:2006,Casini:2007,Hirata:2007}

In $D=1+1$, the connection between entanglement entropy and degree-of-freedom counting is made precise for critical systems
described by a conformal field theory (CFT), for which the entropy of a subregion of size $L$ scales asymptotically as\cite{Holzhey:1994,Calabrese:2004} 
\begin{align}
S = \frac{c}{3} \log{L} + \ldots \label{eqn:1d_scaling}
\end{align}
Here $c$ is the central charge which is a universal number that in a rough sense counts the number of low-energy bosonic
fields---for the case of a CFT described by $N$ free bosonic fields, $c$ is equal to $N$.\cite{Cardy:2010c}
The notion that $c$ measures degrees of freedom is also embodied by the famous $c$-theorem, which states that $c$
decreases monotonically under renormalization group flow.\cite{Zamolodchikov:1986}
Scaling forms for the entanglement entropy containing universal terms analogous to Eq.~(\ref{eqn:1d_scaling}) have been derived or 
conjectured for higher-dimensional critical systems,\cite{Solodukhin:2008,Huerta:2012,Casini:2012,Myers:2012,Lee:2014}
raising the possibility of uncovering new, non-trivial constraints on renormalization flows.  Such an advance
would have broad ramifications for many fields of physics, including the study of quantum critical points
in condensed matter systems.\cite{Grover:2014e}

%-----------------------------
\begin{figure}[t]
\includegraphics[width=0.65\columnwidth]{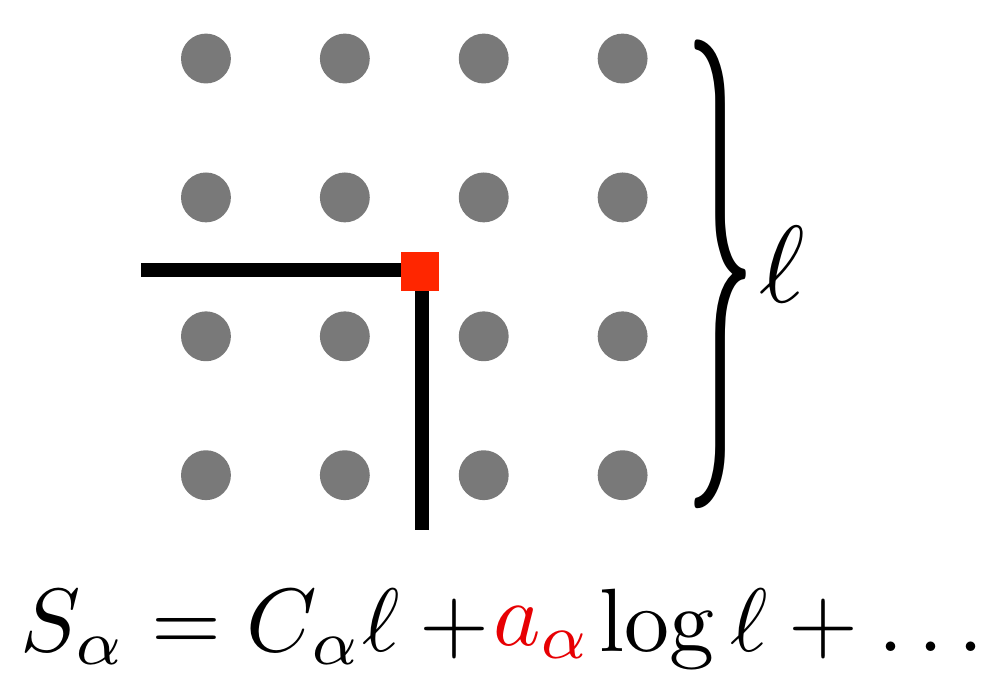}
\caption{For a $D=2+1$ critical system, a sharp corner in the entangling boundary contributes a subleading 
logarithmic term with universal coefficient $a_\alpha$ to the R\'enyi entanglement entropy. }
\label{fig:corner}
\end{figure}
%-----------------------------

In some cases, the relevant entanglement quantities can be calculated directly in continuum field theories\cite{Casini:2007}
and possibly through holographic approaches,\cite{Hirata:2007,Myers:2012,Klebanov:2012} which exploit the 
AdS/CFT correspondence.\cite{Maldacena}  
But in most cases, especially for interacting systems, no exact results are available for the entanglement entropy
at quantum critical points in space-time dimensions of $D=2+1$ or higher.
Thus, in order to make progress in understanding the universal content of the entanglement entropy,
numerical calculations of entanglement need to be performed on a variety of models representing the various strongly-interacting fixed 
points routinely encountered in modern condensed-matter research.

A thriving research effort devoted to such numerical calculations is currently under way.  In order to make concrete comparisons,
practitioners need to reconcile precisely {\it which} entanglement quantities can be simultaneously calculated across lattice models,
continuum field theories, and holographic calculations.  Since the entanglement entropy depends crucially on the geometry of
entangled subregions, this problem largely reduces to a question of subregion shape.  
%Universal log terms appear for smooth bipartitions of D-even systems. \cite{Lee:2014}
%For D-odd systems, log terms arise from singularities in the entangling surface (always?)
Currently, higher-dimensional universal quantities studied in continuum analytical theories typically
arise from smooth curved boundaries,
e.g. for circular subregions.\cite{Casini:2012,CHM_2011,Swingle_2012}
However, problems can arise when trying
to converge subleading scaling terms %for the entanglement entropy 
for such geometries when the theory is regularized on a lattice.\cite{Casini:2009} To date these ``pixelization'' 
issues have not been fully resolved.

Therefore, numerical studies of lattice models in \mbox{$D=2+1$} have largely been restricted to straight boundaries, or boundaries  with 
$\theta=90^{\circ}$ corners, which can be regularized on a square lattice without any artifacts.  
%Although its relationship to monotonicity theorems is unclear,
An interesting universal quantity arises when the entangling boundary contains a sharp corner (or vertex).
For the entanglement entropy $S_\alpha$ with R\'enyi index $\alpha$ (see Eq.~(\ref{RenyiEE})),
a corner with opening angle $\theta$ contributes an additive logarithmic term with a universal coefficient $a_{\alpha}(\theta)$
\begin{equation}
S_{\alpha} = C_\alpha \frac{\ell}{\delta} + a_{\alpha}(\theta) \log \left({ \frac{\ell}{\delta} }\right) + \cdots \label{S_corn}
\end{equation}
Here $\ell$ is the length of the entangling boundary, $\delta$ is the lattice or UV cutoff, $C_\alpha$ is an unknown function of $\alpha$ that 
depends on microscopic details, 
and ellipses represent more rapidly decaying subleading terms and non-universal constants.

The universal ``corner coefficient'' $a_\alpha(\theta)$ has been calculated a number of times in the past, 
in both lattice models and continuum 
field theories.\cite{Fradkin:2006,Casini:2007,Hirata:2007,Singh:2012t,Kallin:2013,Kallin:2014,Helmes:2014,Devakul:2014}
(For an interesting related calculation in \mbox{$D=3+1$} see Ref.~\onlinecite{Devakul:2014e}.)
Strikingly, for $\theta=90^{\circ}$, this quantity appears to not only identify the unique universality class, but also to indicate the underlying 
degrees of freedom of the low-energy theory.
This is already known to be precisely the case for two-dimensional systems constructed out of $(1+1)D$ CFTs (so-called conformal quantum critical points with dynamical exponent $z=2$),
for which the corner coefficient is proportional to the CFT central charge.\cite{Fradkin:2006}
In the Lorentz-invariant ($z=1$) case, series expansion,\cite{Singh:2012t, Devakul:2014} numerical linked cluster\cite{Kallin:2013,Kallin:2014} and quantum Monte Carlo\cite{Inglis_2013,Helmes:2014} studies
of critical lattice models in the $O(N)$ Wilson-Fisher universality class for $N=1$ and $3$
have suggested that the corner coefficient is, to high accuracy, proportional to $N$.
Recently, a series expansion study has examined a bilayer lattice model that can
be continually varied between $N=1,2$ and $3$ critical points, providing additional support for this scenario in an $O(2)$ model.\cite{Devakul:2014}

In this paper, we use a powerful combination of density matrix renormalization group (DMRG) and the numerical linked cluster expansion (NLCE) 
to study two very different lattice models that realize critical points in the $O(2)$ Wilson-Fisher universality class.  For a range of R\'enyi 
indices $\alpha$, the corner-coefficients $a_{\alpha}(\theta=90^{\circ})$ of each lattice model agree to within our numerical confidence, 
offering a striking confirmation of the universality of this quantity.  
Furthermore, our simulations strengthen the evidence that this coefficient is, to remarkably high precision in its leading order, made up of two separate contributions,
\begin{equation}
a_{\alpha}(\theta) \sim N c_{\alpha}(\theta) \label{eqn:conjecture}
\end{equation}
where $N$ is the number of field components in the $O(N)$ theory, and $c_{\alpha}$ is a universal function that appears to be the same for 
all $\phi^4$ theories studied to date. 

%\note{While in principle $a_\alpha$ could have a more complicated dependence on $N$, our results
%below will show that any corrections to the above form must be rather small.}

%found some N dependence in the WF eigenvalues here
%http://www.physics.fsu.edu/Users/Dobrosavljevic/Phase%20Transitions/4-epsilon.pdf

\section{Two-dimensional lattice models with $O(2)$ quantum critical points}

%-----------------------------
\begin{figure}[t]
\includegraphics[width=150px]{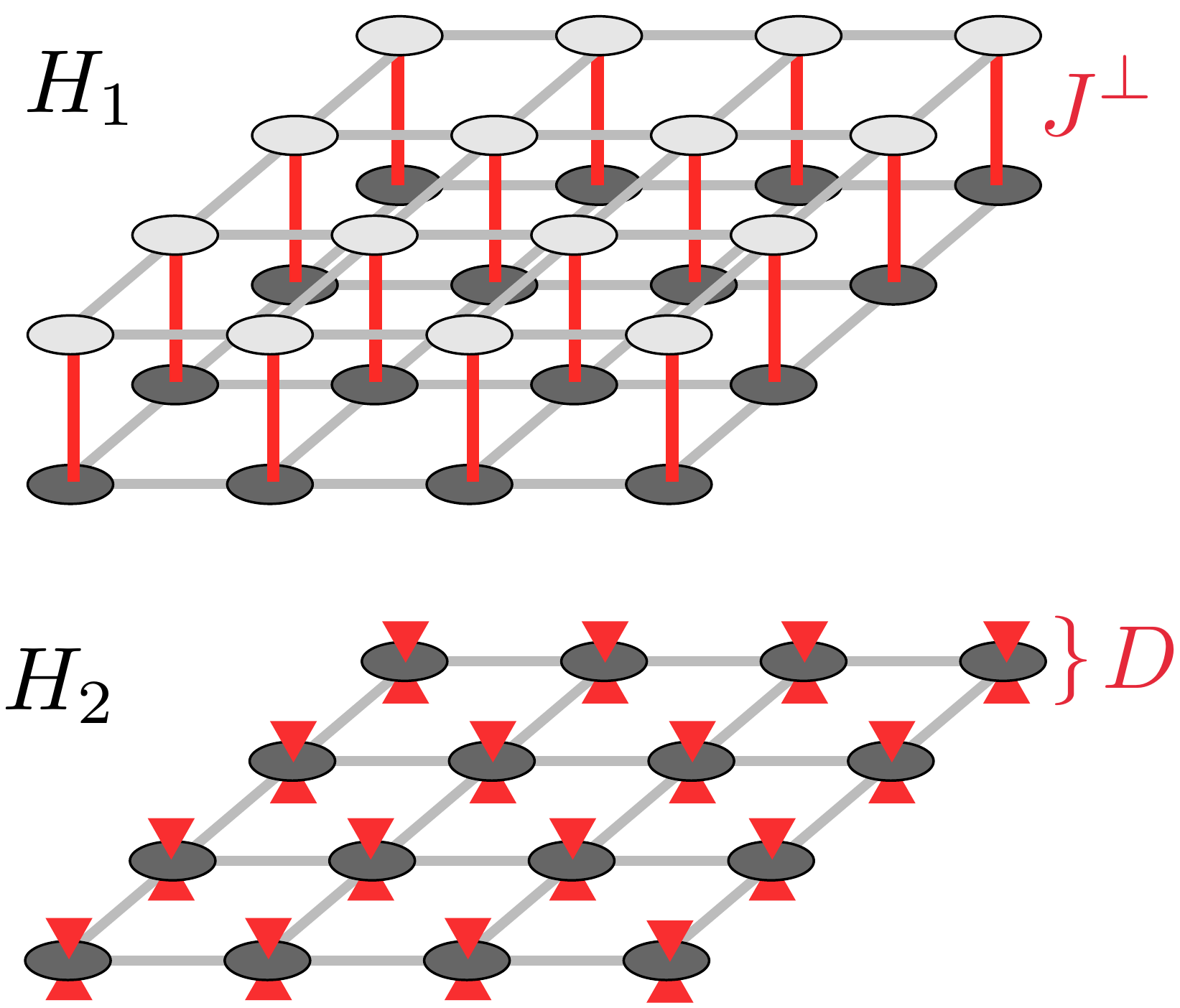}
\caption{Illustrations of the Hamiltonians $H_1$ and $H_2$ \mbox{Eqs.~(\ref{eqn:h1},\ref{eqn:h2})} which both realize a critical point in the $O(2)$ universality class in $(2+1)D$.}
\label{fig:hamiltonians}
\end{figure}
%-----------------------------

Universality is a remarkable phenomenon where asymptotic features of otherwise quite different critical lattice models
are described by a single continuum field theory. This connection is typically demonstrated through critical exponents,
which in many cases can be estimated through perturbative field theory calculations or computed 
numerically using large-scale Monte Carlo calculations.
Instead of looking at critical exponents, in this paper we examine the scaling of the R\'enyi entanglement entropies,
\begin{equation}
S_{\alpha}(A) = \frac{1}{1-\alpha} \log {\rm Tr} (\rho_A^{\alpha}), \label{RenyiEE}
\end{equation}
where $\rho_A = \text{Tr}_B(\rho)$ is the reduced density matrix of a subregion $A$, and $B$ is the rest of the system.
We will calculate $S_{\alpha}$ for geometries with single $\theta=90^{\circ}$ corners for two very different interacting
quantum lattice models, each tuned to a critical point in the $O(2)$ universality class in $(2+1)D$.  These Hamiltonians are:
\begin{align}
H_1 = &  \sum_{n=1,2} \sum_{\langle i, j \rangle} (\sigma^x_{i,n} \sigma^x_{j,n} + \sigma^y_{i,n} \sigma^y_{j,n}) \nonumber \\
        & \mbox{} + J^\perp \sum_i (\sigma^x_{i,1} \sigma^x_{i,2} + \sigma^y_{i,1} \sigma^y_{i,2}) \label{eqn:h1} \\
H_2 = & \sum_{\langle i,j \rangle} \mathbf{S}_i \cdot \mathbf{S}_j + D \sum_i (S^z_i)^2  \label{eqn:h2} \: .
%3.\ H_3 = & \sum_{\langle i, j \rangle} (\sigma^x_i \sigma^x_j + \sigma^y_i \sigma^y_j) + h \sum_i (-1)^{x_i + y_i} \sigma^z_i \label{eqn:h3} \:.
\end{align}

The first model is the spin-$\frac{1}{2}$ XY bilayer where the $\sigma^{x,y,z}$ are the Pauli matrices. This model consists of two stacked layers of the square-lattice XY model, plus a perpendicular XY interaction $J^\perp$ which couples the $i^\text{th}$ site of the top layer to the $i^\text{th}$ site of the bottom layer, as shown in Fig.~\ref{fig:hamiltonians}.

The second model is the $S=1$ square-lattice Heisenberg model plus
single-ion anisotropy $D (S^z_i)^2$. For large $D>0$ the ground state of this model approaches
a product state where every spin is the $m_z=0$ eigenstate of~$S^z$.

%The third model is the $S=\frac{1}{2}$ staggered-field XY model.
%The spins reside on the sites of the two-dimensional square lattice with an applied field in the $z$ direction 
%whose sign alternates between sublattices.

%-----------------------------
\begin{figure}[t]
\includegraphics[width=0.85\columnwidth]{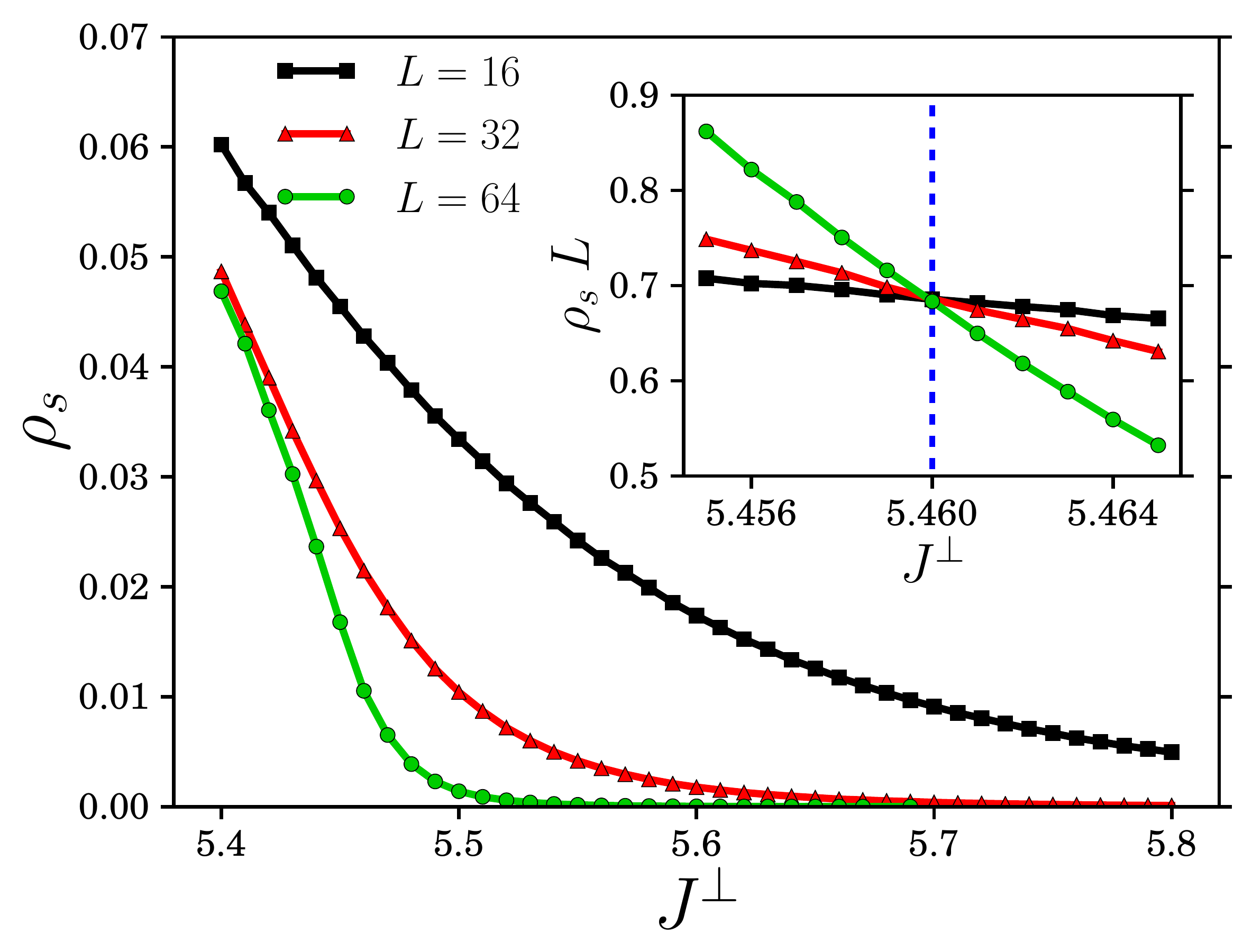}
\caption{Quantum Monte Carlo data for the spin stiffness $\rho_S$ for different linear system sizes $L$, as a 
function of $J^\perp$ for the Hamiltonian $H_1$. The inset shows a scaling plot of $\rho_S L$ to locate the quantum 
critical point, indiated by a vertical dashed line. Errorbars are smaller than the symbol sizes.}
\label{fig:qmc}
\end{figure}
%-----------------------------

Unless noted otherwise, we fix each of the above models to specific quantum critical points
corresponding to the following parameter values:
\begin{align}
1.\ & J^\perp_c  = 5.460(1) \\
2.\ & D_c  = 5.625(5) \: .
%3.\ & h_c  = 1.99(1) \ .
\end{align}
While $D_c$ was taken from Ref.~\onlinecite{Zhang:2013p}, we employed stochastic series expansion quantum 
Monte Carlo simulations~\cite{Sandvik:1999} to locate $J^\perp_c$. For this purpose, we considered the spin stiffness 
$\rho_S$, obtained as 
$\rho_S=\langle \mathbf{W}^2 \rangle /(2\beta)$
from the winding number 
fluctuations.\cite{Pollock_Ceperley} Here, $\beta$ denotes the inverse temperature, scaled in the simulations as $\beta=4 L$ with the 
linear system size $L$ in order to probe quantum critical scaling properties. 
Fig.~\ref{fig:qmc} shows $\rho_S$ as a function of $J^\perp$ for different system sizes near criticality.  
At the $O(2)$ quantum critical point in $(2+1)D$, $\rho_S$ scales proportional to $1/L$, 
and we extract $J^\perp_c$ from the data crossing shown in the inset of Fig.~\ref{fig:qmc}.

Both critical points are transitions between a phase with spontaneous antiferromagnetic XY 
order and a trivial phase in which all Hamiltonian symmetries are restored.
Thus these critical points belong to the interacting $O(2)$ universality class in $(2+1)D$, and offer
an excellent testbed for confirming the universality of the corner coefficient, the calculation 
of which we now discuss.

\section{Method for Computing the Corner Coefficient}

%%-----------------------------
%\begin{figure}[t]
%\includegraphics[width=\columnwidth]{corner.pdf}
%\caption{Isolating the entanglement corner contribution for a specific plaquette of a square-lattice cluster.}
%\label{fig:corner}
%\end{figure}
%%-----------------------------

The numerical calculation of universal terms in the scaling of R\'enyi entanglement entropies poses two main challenges.
First, the ability to reliably access $S_{\alpha}(A)$ in a general way can be difficult, even for modern numerical methods.
For example, Monte Carlo or series expansion methods can access only integer values of $\alpha \ge 2$;
on the other hand, the cost of DMRG can rise sharply depending on the geometry of region $A$.
Second, even if a reliable estimator for $S_{\alpha}(A)$ is obtained for a given method,
the desired universal scaling terms are generally sub-leading in spatial dimensions higher than one, such as in
Eq.~(\ref{S_corn}).
This means that numerical signals may easily be overwhelmed with ``noise'' coming from the leading-order 
(non-universal) area-law term.  
As demonstrated in previous works,\cite{Kallin:2013,Kallin:2014} the powerful numerical linked-cluster expansion (NLCE) offers a way to 
combat this challenge, using relatively moderate computing resources. 
%when compared to other finite-size lattice methods.

The NLCE is a method for computing thermodynamic properties
from a series of numerical calculations of finite clusters on a lattice.\cite{Rigol:2006,Rigol:2007_1,Rigol:2007_2,Tang:2013,Kallin:2013,Kallin:2014}
In its original formulation, which includes all connected clusters up to a given size, 
one encounters a bottleneck arising from the task of computing every possible way these clusters can be embedded on a lattice;
this severely limits the maximum cluster size (to about 16 sites).\cite{Tang:2013} 
Here we follow the approach of Ref.~\onlinecite{Kallin:2013} and consider only rectangular
clusters, relevant for the calculation of quantities on square-lattice systems.
This approach has been demonstrated to give excellent convergence, and makes the embedding problem simple
enough that the only remaining limitation becomes the maximum cluster size reachable by the numerical solver.

Most previous NLCE studies have used Lanczos exact diagonalization as the cluster solver,
but in principle any numerical method can be used.
Here we choose the density matrix renormalization group (DMRG) method,\cite{Schollwoeck:2005,ITensor}
adapted to work on 2D clusters.\cite{Stoudenmire}
Unlike Lanczos which scales exponentially in the number of lattice sites of a two-dimensional cluster,
DMRG has a cost scaling exponentially with only the linear size of the cluster 
(more precisely, only the smaller of the two linear sizes, but for the entanglement corner coefficient discussed below we compute both orientations).
Computing entanglement entropy is very efficient and simple within DMRG because the spectrum of the reduced density matrix 
of various bipartitions is automatically computed as part of the DMRG algorithm.
The only difficulty is that, to compute the entanglement of some given region A, 
the one-dimensional path used internally within DMRG must pass through \emph{all} the sites of region~A 
before visiting the other sites of the system. Thus, for each cluster in the NLCE, we performed separate DMRG calculations 
with different paths in order to obtain every entanglement cut required for computing the corner terms.

Next, to perform the NLCE using the DMRG solver, the calculation has to be organized into clusters and sub-clusters, 
each defined by the maximum desired ``order'' of the NLCE calculation (see below).
For each cluster, we are interested in isolating {\it only} the entanglement due to a
$\theta = 90^{\circ}$ corner, which is a sub-leading correction to the area-law scaling as in Eq.~(\ref{S_corn}).
Unlike other methods, such as direct calculation of square subregions in a toroidal lattice,\cite{Inglis_2013}
the NLCE offers an advantage that this sub-leading contribution can be isolated for each cluster individually.
This procedure, detailed in Ref.~\onlinecite{Kallin:2014}, involves
adding the entanglement for two cuts which have identical corner contributions and complementary line contributions,
then subtracting off the line contributions computed from separate straight-line cuts.
Because the NLCE is designed to compute extensive properties, we define the ``property'' of a cluster used
in the expansion\cite{Kallin:2014} to be the {\it sum} of the corner coefficient of every plaquette of the cluster.
%For each element of this sum, it thus requires up to four individual DMRG calculations per cluster to extract the  corner contribution.

%As mentioned above, the NLCE 

%To isolate the corner contribution to the entanglement for a $90^\circ$ corner located in a specific plaquette
%as in Fig.~\ref{fig:corner}(a), we use the procedure shown in Fig.~\ref{fig:corner}(b). 
%One adds the entanglement for two cuts which have identical corner contributions and complementary line contributions,
%then subtracts off the line contributions computed from straight-line cuts.
%Because the NLCE is designed to compute extensive properties, we define the property $P_c$ of the cluster $c$ used
%in the expansion to be the sum of the corner coefficient of every plaquette of the cluster.

The NLCE was originally conceived for properties which converge to a finite value in the thermodynamic limit. 
For such properties, convergence is heuristically reached when the maximum cluster size of the NLCE surpasses 
the finite correlation length.
In contrast, the critical systems we study here have an infinite correlation length;
also the corner contribution to the entanglement is made up of contributions from all length scales,
thus it diverges with increasing cluster size.
Specifically, by removing the leading-order piece of Eq.~(\ref{S_corn}), the corner term is 
expected to diverge as
\begin{equation}
\mathcal{V}_\alpha = a_\alpha \log \ell + b_\alpha, \label{eqn:log}
\end{equation}
where $\ell$ a length-scale associated with the size of region $A$, and non-universal constants are
combined into $b_\alpha$.
In the NLCE, this length-scale is related in some way to the maximum cluster size employed
in the expansion.  In this paper, we equate $\ell$ to the
linear size of the maximum cluster, %s included in the NLCE,
defined as $\ell=\frac{1}{2}(N_x + N_y)$ where $N_x$, $N_y$ are the rectangular cluster dimensions.
We call this definition of $\ell$ ``arithmetic'' order---note however that other definitions of $\ell$ are possible.\cite{Kallin:2014,Kallin_thesis}
In other words, to define the length-scale $\ell$, we terminate the NLCE at different maximum orders, 
meaning that we include only clusters of linear size less than or equal to $\ell$.
Finally, we extract the corner coefficient $a_\alpha$ from the slope of a linear fit of the data as a function of $\log \ell$.
This procedure is illustrated in the next section for the specific models considered in this paper.

%-----------------------------
\begin{figure}[t]
\includegraphics[width=0.85\columnwidth]{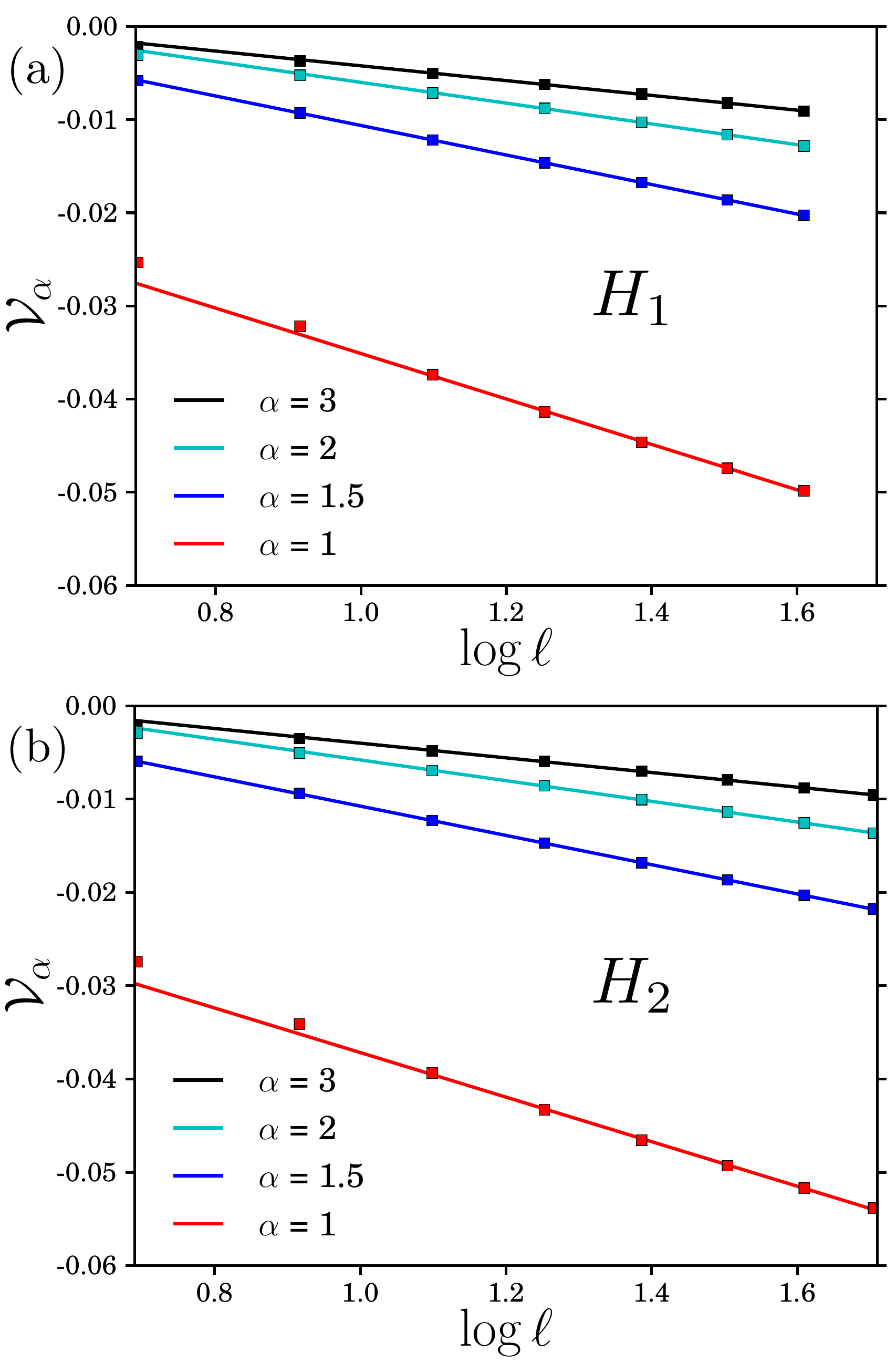}
\caption{Linear fits of the corner contribution to the entanglement as a function the logarithm of the maximum cluster size 
$\ell$ for selected Renyi indices~$\alpha$. The data for orders $\ell = 2.0$ and $\ell=2.5$ was discarded when computing the fits.
Panel (a) shows results for the XY bilayer model $H_1$ and panel (b) the $S=1$ model $H_2$.}
\label{fig:fits}
\end{figure}
%-----------------------------

\section{Results}

We turn now to a detailed calculation of the corner coefficient using NLCE for two strongly-interacting
quantum lattice models.
First, for the XY bilayer model $H_1$, we performed the calculation for cluster sizes up to and including arithmetic order $\ell=5.0$; 
that is, the largest clusters solved with DMRG had dimensions $(N_x+N_y)/2=5.0$.
For these bilayer clusters, $N_x$ and $N_y$ refer to the number of sites in a single layer,
%the $x$ and $y$ directions in a ``top-down'' view as if the system was a single layer---
so that the total number of lattice sites per cluster is $2 N_x N_y$.
For the anisotropic $S=1$ Heisenberg model $H_2$, we summed clusters through order $\ell=5.5$.
In this case, $N_x N_y$ is the total number of lattice sites.
For both systems, by keeping up to $m=10,000$ states, the 
DMRG calculations for the clusters reached truncation errors of no more than $5\times 10^{-10}$ and typically
much smaller, meaning the DMRG results were essentially exact.

Figure \ref{fig:fits} shows the NLCE results for the corner contributions \mbox{$\mathcal{V}_\alpha$}
at each order for selected values of the Renyi index $\alpha$. 
In performing the fits we discarded the lowest two orders $\ell = 2.0$ and $\ell = 2.5$. 
As can be seen from the figures, for linear sizes $\ell \geq 3.0$, the data already shows excellent agreement with
the asymptotic form of Eq.~(\ref{eqn:log}) for both models.
The slopes of the linear fits are our estimates of the corner coefficient $a_\alpha$.

%-----------------------------
\begin{figure}[t]
\includegraphics[width=0.85\columnwidth]{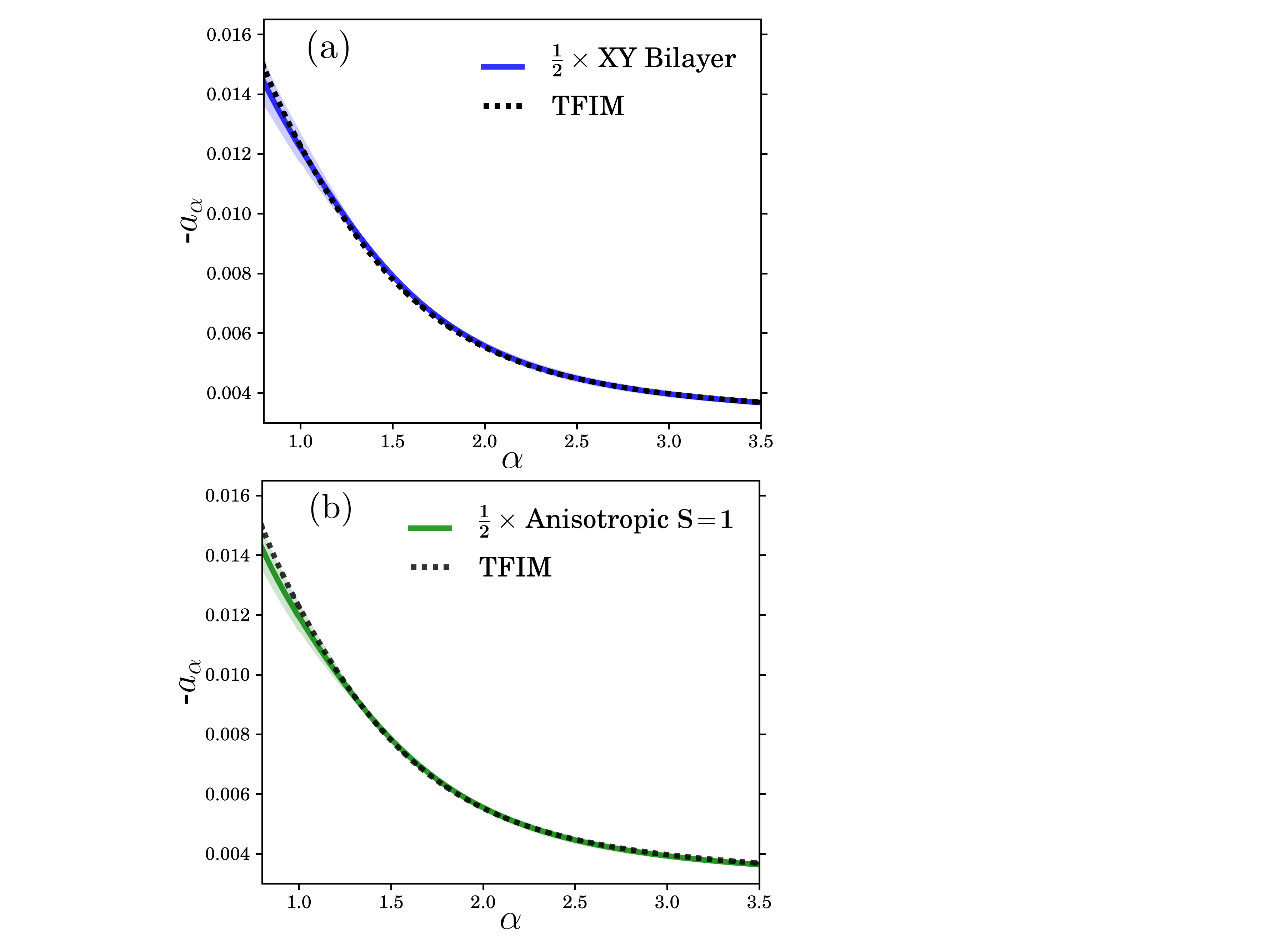}
\caption{Corner coefficients $a_\alpha$ for (a) the XY bilayer model $H_1$ and (b) the anisotropic $S=1$ model $H_2$. 
Because the coefficients are negative, we plot $-a_\alpha$ and also divide the results of the $O(2)$ models by two for comparison
to the $O(1)$ transverse-field Ising model (TFIM) results.}
\label{fig:tfimcomparisons}
\end{figure}
%-----------------------------

Figure \ref{fig:tfimcomparisons} shows the resulting corner coefficients $a_\alpha$ thus extracted for the two models,
$H_1$ and $H_2$, at their respective critical points.
Since, as discussed above, these quantum critical points share an O(2) universality class,
we address the conjecture posed in Eq.~(\ref{eqn:conjecture}) by dividing each by a factor of $N=2$, and compare
to a calculation performed on the transverse-field Ising model (TFIM) described by a scalar ($N=1$) $\phi^4$ theory. 
As illustrated in Fig.~\ref{fig:tfimcomparisons}, 
the coefficients of both $O(2)$ models are in excellent agreement with each other and,
within error bars, twice that of the TFIM.
Note that the TFIM results shown here are computed following the NLCE procedure discussed in Ref.~\onlinecite{Kallin:2013},
however new data for larger cluster sizes (solved using DMRG) is included up to arithmetic order $\ell=6.0$.
This allows us to apply exactly
the same fitting procedure for the TFIM as for the $O(2)$ models presently studied,
and thus to make a direct comparison.
To account for any uncertainty in fitting procedure used for the $O(2)$ systems (Fig.~\ref{fig:fits}),
the light shaded regions in each plot of Fig.~\ref{fig:tfimcomparisons} shows the difference in the corner coefficient that 
would result from fitting the NLCE corner term data only for orders $\ell \geq 4.0$.

\begin{figure}[t]
\includegraphics[width=0.85\columnwidth]{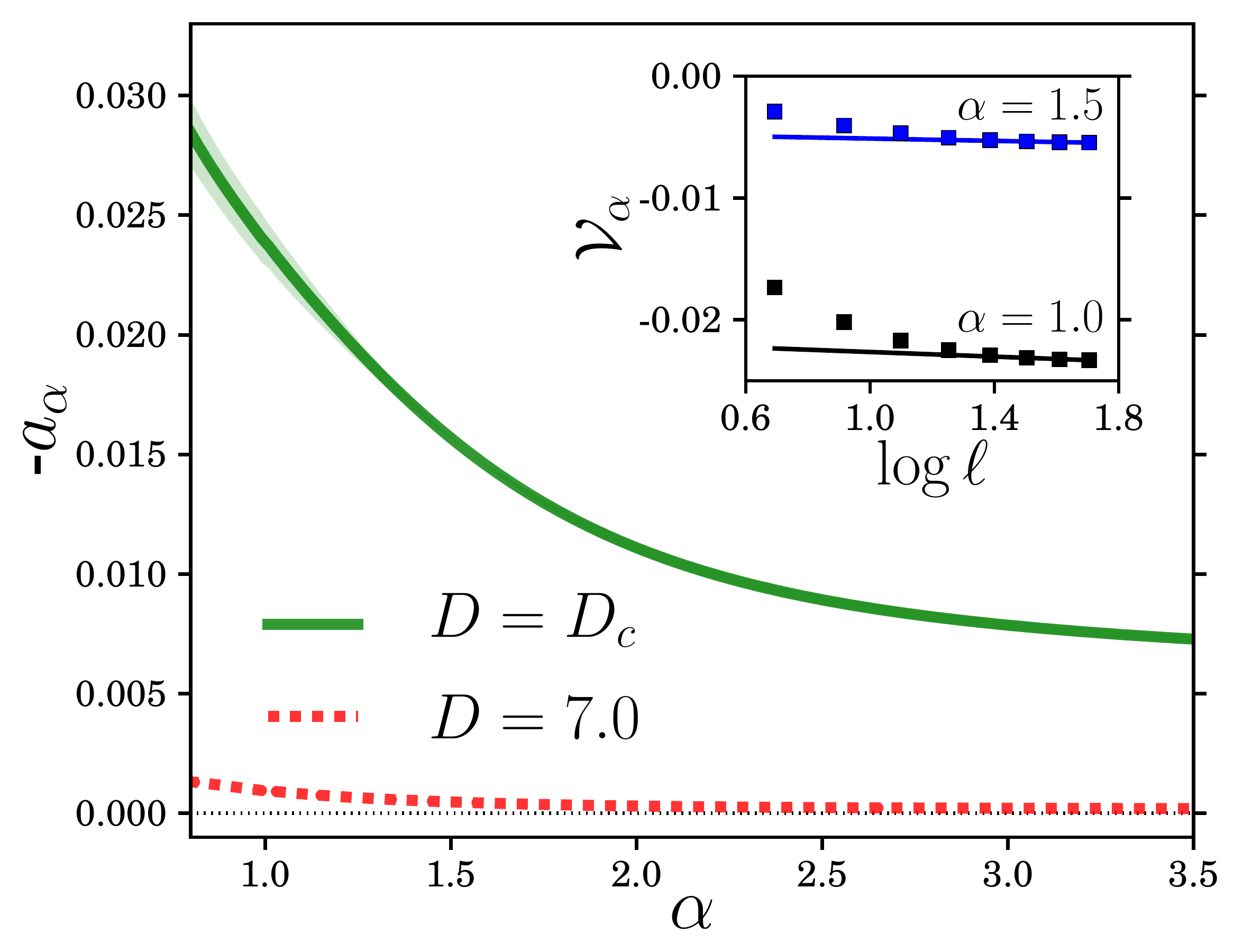}
\caption{Comparison of the corner coefficient of the anisotropic $S=1$ model at the critical point $D_c=5.625$ (upper curve) with the 
corner coefficient at $D=7$ (lower curve). The $D=7$ coefficient is consistent with $a_\alpha=0$ as expected for a trivial gapped disordered phase. 
The inset shows the linear fits used to compute $a_\alpha$ for the lower curve for $\alpha=1.0$ and $1.5$.}
\label{fig:d7s1compare}
\end{figure}

Finally, we note that with the adaptation of DMRG as a cluster solver for the R\'enyi entropy,
the quality of the NLCE extrapolation of the corner term is approaching an accuracy sufficient to 
distinguish the critical regime of the model from non-critical regimes.  
As a demonstration, we carry out a calculation of the corner coefficient of the $S=1$ system $H_2$ 
with the anisotropy increased from the critical value of \mbox{$D_c = 5.625$} to a much larger value, \mbox{$D=7$}.
For \mbox{$D>D_c$} the system is in a trivial gapped phase, for which the corner coefficient is expected to be 
\mbox{$a_\alpha = 0$} for all $\alpha$. 
Figure~\ref{fig:d7s1compare} shows that the NLCE results are consistent with this expectation.
For the \mbox{$D=7$} model the largest clusters have linear size $\ell=5.5$.
As shown in the inset of Fig.~\ref{fig:d7s1compare}, the logarithmic scaling ansatz of 
Eq.~(\ref{eqn:log}) breaks down, so that the data fits very accurately to the form,
\begin{equation}
\mathcal{V}_\alpha = f_\alpha e^{-\ell / \xi_\alpha} + b_\alpha,
\end{equation} 
as might be expected for a system with a finite correlation length (exponential fits are not shown).
%where $f_\alpha$, $\xi_\alpha$, and $b_\alpha$ are fitting parameters (exponential fits not shown).
Continuing to use linear fits versus $\log \ell$, but only for the largest-order NLCE data $\ell \geq 4.0$,
the resulting slopes plotted in Fig.~\ref{fig:d7s1compare} show no evidence for any logarithmic scaling.

%-----------------------------
%-----------------------------

\section{Discussion}

In this paper, we have studied
the scaling of the R\'enyi entanglement entropies of
two very different strongly-interacting lattice models in two spatial dimensions,
using a combination of the density matrix renormalization group (DMRG) and a numerical linked cluster expansion
(NLCE) procedure.
Each model is separately tuned to its respective quantum critical points, which both lie in the 
$D=2+1$ dimensional $O(N)$ Wilson-Fisher universality class with $N=2$.
By isolating the contribution to the entanglement entropy scaling due to the presence of a $90^\circ$
corner in the boundary between entangled subregions, we show the presence of a clear sub-leading 
additive logarithmic scaling term $a_{\alpha}$
The value of the coefficient of this logarithmic term is identical to within numerical precision for the 
two different lattice models, which provides a striking demonstration of the universality of this quantity
for a wide range of R\'enyi indices $\alpha$.

Further, the value obtained for $a_{\alpha}$ in both models is, to within numerical precision, 
twice as large as the same coefficient for the Ising ($N=1$) limit of the Wilson-Fisher universality class, 
which can be calculated using the same numerical procedure at the quantum critical point of a 
transverse field Ising model.
Together with recent numerical results for bilayers realizing $O(2)$ and $O(3)$ critical behavior,\cite{Kallin:2014,Devakul:2014,Helmes:2014} 
analytical results for conformal critical points,\cite{Fradkin:2006} and free field theory calculations,\cite{Casini:2007} 
there is a growing body of evidence that this universal corner coefficient 
reflects, in its leading-order behavior, the low-energy degrees of freedom of the associated strongly-interacting critical field theory.
Based on the high numerical accuracy of this leading order term (Eq.~\ref{eqn:conjecture}), it would be interesting to
examine field theory calculations at the $O(N)$ Wilson-Fisher fixed point to see if this behavior is reproduced.

Our result could introduce a powerful new tool in the arsenal of condensed-matter and quantum field theorists
by providing a simple universal quantity that not only can distinguish between different universality classes, but
can elucidate the low-energy structure of the quantum critical theory in a practical calculation.
It remains to be seen how the corner coefficient behaves for other strongly interacting classes of critical
systems that, unlike the $O(N)$ Wilson-Fisher fixed point, are not perturbatively close (in an $\epsilon$-expansion sense) 
to a non-interacting fixed point.
To clarify this situation, it will be interesting to apply the numerical framework used here to richer
examples of interacting $D=2+1$ critical points.  
Of immediate interest is the deconfined quantum critical point observed in the square-lattice \mbox{$J$--$Q$} model\cite{Senthil:2004,Sandvik:2010d}
and larger-$N$ extensions,\cite{Block_2013} which may lie in the (non-compact) CP$^{N-1}$ universality class.

\subsection*{Acknowledgments} 
We would like to acknowledge crucial discussions with D.~Chowdhury, T.~Grover, A.~Kallin, C.~Pashartis, M.~Metlitski, 
R.~Myers,  R.~Singh, B.~Swingle, and W.~Witczak-Krempa.
We would like to especially thank J.~Carrasquilla for recommending the anisotropic \mbox{$S=1$} model
as an example of O(2) critical behavior.
The simulations were performed on the computing facilities of SHARCNET and on the Perimeter Institute HPC.  Support was provided by NSERC, the Canada Research Chair program, the Ontario Ministry of Research and Innovation, the John Templeton Foundation, and the Perimeter Institute (PI) for Theoretical Physics. Research at PI is supported by the Government of Canada through Industry Canada and by the Province of Ontario through the Ministry of Economic Development \& Innovation.  
Financial support by the DFG under Grant WE 3649/3-1 is gratefully acknowledged, as well as the allocation of CPU time within JARA-HPC at RWTH Aachen University and JSC J\"ulich.

\bibliography{o2_dmrg_nlce}

\begin{thebibliography}{39}%
\makeatletter
\providecommand \@ifxundefined [1]{%
 \@ifx{#1\undefined}
}%
\providecommand \@ifnum [1]{%
 \ifnum #1\expandafter \@firstoftwo
 \else \expandafter \@secondoftwo
 \fi
}%
\providecommand \@ifx [1]{%
 \ifx #1\expandafter \@firstoftwo
 \else \expandafter \@secondoftwo
 \fi
}%
\providecommand \natexlab [1]{#1}%
\providecommand \enquote  [1]{``#1''}%
\providecommand \bibnamefont  [1]{#1}%
\providecommand \bibfnamefont [1]{#1}%
\providecommand \citenamefont [1]{#1}%
\providecommand \href@noop [0]{\@secondoftwo}%
\providecommand \href [0]{\begingroup \@sanitize@url \@href}%
\providecommand \@href[1]{\@@startlink{#1}\@@href}%
\providecommand \@@href[1]{\endgroup#1\@@endlink}%
\providecommand \@sanitize@url [0]{\catcode `\\12\catcode `\$12\catcode
  `\&12\catcode `\#12\catcode `\^12\catcode `\_12\catcode `\%12\relax}%
\providecommand \@@startlink[1]{}%
\providecommand \@@endlink[0]{}%
\providecommand \url  [0]{\begingroup\@sanitize@url \@url }%
\providecommand \@url [1]{\endgroup\@href {#1}{\urlprefix }}%
\providecommand \urlprefix  [0]{URL }%
\providecommand \Eprint [0]{\href }%
\providecommand \doibase [0]{http://dx.doi.org/}%
\providecommand \selectlanguage [0]{\@gobble}%
\providecommand \bibinfo  [0]{\@secondoftwo}%
\providecommand \bibfield  [0]{\@secondoftwo}%
\providecommand \translation [1]{[#1]}%
\providecommand \BibitemOpen [0]{}%
\providecommand \bibitemStop [0]{}%
\providecommand \bibitemNoStop [0]{.\EOS\space}%
\providecommand \EOS [0]{\spacefactor3000\relax}%
\providecommand \BibitemShut  [1]{\csname bibitem#1\endcsname}%
\let\auto@bib@innerbib\@empty
%</preamble>
\bibitem [{\citenamefont {Fradkin}\ and\ \citenamefont
  {Moore}(2006)}]{Fradkin:2006}%
  \BibitemOpen
  \bibfield  {author} {\bibinfo {author} {\bibfnamefont {E.}~\bibnamefont
  {Fradkin}}\ and\ \bibinfo {author} {\bibfnamefont {J.~E.}\ \bibnamefont
  {Moore}},\ }\href@noop {} {\bibfield  {journal} {\bibinfo  {journal} {Phys.
  Rev. Lett.}\ }\textbf {\bibinfo {volume} {97}},\ \bibinfo {pages} {050404}
  (\bibinfo {year} {2006})}\BibitemShut {NoStop}%
\bibitem [{\citenamefont {Casini}\ and\ \citenamefont
  {Huerta}(2007)}]{Casini:2007}%
  \BibitemOpen
  \bibfield  {author} {\bibinfo {author} {\bibfnamefont {H.}~\bibnamefont
  {Casini}}\ and\ \bibinfo {author} {\bibfnamefont {M.}~\bibnamefont
  {Huerta}},\ }\href@noop {} {\bibfield  {journal} {\bibinfo  {journal}
  {Nuclear Physics B}\ }\textbf {\bibinfo {volume} {764}},\ \bibinfo {pages}
  {183 } (\bibinfo {year} {2007})}\BibitemShut {NoStop}%
\bibitem [{\citenamefont {Hirata}\ and\ \citenamefont
  {Takayanagi}(2007)}]{Hirata:2007}%
  \BibitemOpen
  \bibfield  {author} {\bibinfo {author} {\bibfnamefont {T.}~\bibnamefont
  {Hirata}}\ and\ \bibinfo {author} {\bibfnamefont {T.}~\bibnamefont
  {Takayanagi}},\ }\href@noop {} {\bibfield  {journal} {\bibinfo  {journal}
  {Journal of High Energy Physics}\ }\textbf {\bibinfo {volume} {2007}},\
  \bibinfo {pages} {042} (\bibinfo {year} {2007})}\BibitemShut {NoStop}%
\bibitem [{\citenamefont {Holzhey}\ \emph {et~al.}(1994)\citenamefont
  {Holzhey}, \citenamefont {Larsen},\ and\ \citenamefont
  {Wilczek}}]{Holzhey:1994}%
  \BibitemOpen
  \bibfield  {author} {\bibinfo {author} {\bibfnamefont {C.}~\bibnamefont
  {Holzhey}}, \bibinfo {author} {\bibfnamefont {F.}~\bibnamefont {Larsen}}, \
  and\ \bibinfo {author} {\bibfnamefont {F.}~\bibnamefont {Wilczek}},\
  }\href@noop {} {\bibfield  {journal} {\bibinfo  {journal} {Nucl. Phys. B}\
  }\textbf {\bibinfo {volume} {424}},\ \bibinfo {pages} {443} (\bibinfo {year}
  {1994})}\BibitemShut {NoStop}%
\bibitem [{\citenamefont {Calabrese}\ and\ \citenamefont
  {Cardy}(2004)}]{Calabrese:2004}%
  \BibitemOpen
  \bibfield  {author} {\bibinfo {author} {\bibfnamefont {P.}~\bibnamefont
  {Calabrese}}\ and\ \bibinfo {author} {\bibfnamefont {J.}~\bibnamefont
  {Cardy}},\ }\href@noop {} {\bibfield  {journal} {\bibinfo  {journal} {Journal
  of Statistical Mechanics: Theory and Experiment}\ }\textbf {\bibinfo {volume}
  {2004}},\ \bibinfo {pages} {P06002} (\bibinfo {year} {2004})}\BibitemShut
  {NoStop}%
\bibitem [{\citenamefont {Cardy}(2010)}]{Cardy:2010c}%
  \BibitemOpen
  \bibfield  {author} {\bibinfo {author} {\bibfnamefont {J.}~\bibnamefont
  {Cardy}},\ }\href@noop {} {\bibfield  {journal} {\bibinfo  {journal} {Journal
  of Statistical Mechanics: Theory and Experiment}\ }\textbf {\bibinfo {volume}
  {2010}},\ \bibinfo {pages} {P10004} (\bibinfo {year} {2010})}\BibitemShut
  {NoStop}%
\bibitem [{\citenamefont {Zamolodchikov}(1986)}]{Zamolodchikov:1986}%
  \BibitemOpen
  \bibfield  {author} {\bibinfo {author} {\bibfnamefont {A.}~\bibnamefont
  {Zamolodchikov}},\ }\href@noop {} {\bibfield  {journal} {\bibinfo  {journal}
  {JETP Lett.}\ }\textbf {\bibinfo {volume} {43}},\ \bibinfo {pages} {730}
  (\bibinfo {year} {1986})}\BibitemShut {NoStop}%
\bibitem [{\citenamefont {Solodukhin}(2008)}]{Solodukhin:2008}%
  \BibitemOpen
  \bibfield  {author} {\bibinfo {author} {\bibfnamefont {S.~N.}\ \bibnamefont
  {Solodukhin}},\ }\href {\doibase
  http://dx.doi.org/10.1016/j.physletb.2008.05.071} {\bibfield  {journal}
  {\bibinfo  {journal} {Physics Letters B}\ }\textbf {\bibinfo {volume}
  {665}},\ \bibinfo {pages} {305 } (\bibinfo {year} {2008})}\BibitemShut
  {NoStop}%
\bibitem [{\citenamefont {Huerta}(2012)}]{Huerta:2012}%
  \BibitemOpen
  \bibfield  {author} {\bibinfo {author} {\bibfnamefont {M.}~\bibnamefont
  {Huerta}},\ }\href {\doibase
  http://dx.doi.org/10.1016/j.physletb.2012.03.044} {\bibfield  {journal}
  {\bibinfo  {journal} {Physics Letters B}\ }\textbf {\bibinfo {volume}
  {710}},\ \bibinfo {pages} {691 } (\bibinfo {year} {2012})}\BibitemShut
  {NoStop}%
\bibitem [{\citenamefont {Casini}\ and\ \citenamefont
  {Huerta}(2012)}]{Casini:2012}%
  \BibitemOpen
  \bibfield  {author} {\bibinfo {author} {\bibfnamefont {H.}~\bibnamefont
  {Casini}}\ and\ \bibinfo {author} {\bibfnamefont {M.}~\bibnamefont
  {Huerta}},\ }\href@noop {} {\bibfield  {journal} {\bibinfo  {journal} {Phys.
  Rev. D}\ }\textbf {\bibinfo {volume} {85}},\ \bibinfo {pages} {125016}
  (\bibinfo {year} {2012})}\BibitemShut {NoStop}%
\bibitem [{\citenamefont {Myers}\ and\ \citenamefont
  {Singh}(2012)}]{Myers:2012}%
  \BibitemOpen
  \bibfield  {author} {\bibinfo {author} {\bibfnamefont {R.~C.}\ \bibnamefont
  {Myers}}\ and\ \bibinfo {author} {\bibfnamefont {A.}~\bibnamefont {Singh}},\
  }\href@noop {} {\bibfield  {journal} {\bibinfo  {journal} {Journal of High
  Energy Physics}\ }\textbf {\bibinfo {volume} {2012}},\ \bibinfo {pages} {1}
  (\bibinfo {year} {2012})}\BibitemShut {NoStop}%
\bibitem [{\citenamefont {Lee}\ \emph {et~al.}(2014)\citenamefont {Lee},
  \citenamefont {McGough},\ and\ \citenamefont {Safdi}}]{Lee:2014}%
  \BibitemOpen
  \bibfield  {author} {\bibinfo {author} {\bibfnamefont {J.}~\bibnamefont
  {Lee}}, \bibinfo {author} {\bibfnamefont {L.}~\bibnamefont {McGough}}, \ and\
  \bibinfo {author} {\bibfnamefont {B.~R.}\ \bibnamefont {Safdi}},\ }\href@noop
  {} {\bibfield  {journal} {\bibinfo  {journal} {Phys. Rev. D}\ }\textbf
  {\bibinfo {volume} {89}},\ \bibinfo {pages} {125016} (\bibinfo {year}
  {2014})}\BibitemShut {NoStop}%
\bibitem [{\citenamefont {Grover}(2014)}]{Grover:2014e}%
  \BibitemOpen
  \bibfield  {author} {\bibinfo {author} {\bibfnamefont {T.}~\bibnamefont
  {Grover}},\ }\href@noop {} {\bibfield  {journal} {\bibinfo  {journal} {Phys.
  Rev. Lett.}\ }\textbf {\bibinfo {volume} {112}},\ \bibinfo {pages} {151601}
  (\bibinfo {year} {2014})}\BibitemShut {NoStop}%
\bibitem [{\citenamefont {Klebanov}\ \emph {et~al.}(2012)\citenamefont
  {Klebanov}, \citenamefont {Nishioka}, \citenamefont {Pufu},\ and\
  \citenamefont {Safdi}}]{Klebanov:2012}%
  \BibitemOpen
  \bibfield  {author} {\bibinfo {author} {\bibfnamefont {I.~R.}\ \bibnamefont
  {Klebanov}}, \bibinfo {author} {\bibfnamefont {T.}~\bibnamefont {Nishioka}},
  \bibinfo {author} {\bibfnamefont {S.~S.}\ \bibnamefont {Pufu}}, \ and\
  \bibinfo {author} {\bibfnamefont {B.~R.}\ \bibnamefont {Safdi}},\ }\href@noop
  {} {\bibfield  {journal} {\bibinfo  {journal} {Journal of High Energy
  Physics}\ }\textbf {\bibinfo {volume} {2012}} (\bibinfo {year}
  {2012})}\BibitemShut {NoStop}%
\bibitem [{\citenamefont {Maldacena}(1998)}]{Maldacena}%
  \BibitemOpen
  \bibfield  {author} {\bibinfo {author} {\bibfnamefont {J.~M.}\ \bibnamefont
  {Maldacena}},\ }\href@noop {} {\bibfield  {journal} {\bibinfo  {journal}
  {Adv. Theor. Math. Phys.}\ }\textbf {\bibinfo {volume} {2}},\ \bibinfo
  {pages} {231} (\bibinfo {year} {1998})}\BibitemShut {NoStop}%
\bibitem [{\citenamefont {Casini}\ \emph {et~al.}(2011)\citenamefont {Casini},
  \citenamefont {Huerta},\ and\ \citenamefont {Myers}}]{CHM_2011}%
  \BibitemOpen
  \bibfield  {author} {\bibinfo {author} {\bibfnamefont {H.}~\bibnamefont
  {Casini}}, \bibinfo {author} {\bibfnamefont {M.}~\bibnamefont {Huerta}}, \
  and\ \bibinfo {author} {\bibfnamefont {R.}~\bibnamefont {Myers}},\ }\href
  {http://dx.doi.org/10.1007/JHEP05%282011%29036} {\bibfield  {journal}
  {\bibinfo  {journal} {Journal of High Energy Physics}\ }\textbf {\bibinfo
  {volume} {2011}},\ \bibinfo {eid} {36} (\bibinfo {year} {2011})}\BibitemShut
  {NoStop}%
\bibitem [{\citenamefont {Swingle}\ and\ \citenamefont
  {Senthil}(2012)}]{Swingle_2012}%
  \BibitemOpen
  \bibfield  {author} {\bibinfo {author} {\bibfnamefont {B.}~\bibnamefont
  {Swingle}}\ and\ \bibinfo {author} {\bibfnamefont {T.}~\bibnamefont
  {Senthil}},\ }\href@noop {} {\bibfield  {journal} {\bibinfo  {journal} {Phys.
  Rev. B}\ }\textbf {\bibinfo {volume} {86}},\ \bibinfo {pages} {155131}
  (\bibinfo {year} {2012})}\BibitemShut {NoStop}%
\bibitem [{\citenamefont {Casini}\ and\ \citenamefont
  {Huerta}(2009)}]{Casini:2009}%
  \BibitemOpen
  \bibfield  {author} {\bibinfo {author} {\bibfnamefont {H.}~\bibnamefont
  {Casini}}\ and\ \bibinfo {author} {\bibfnamefont {M.}~\bibnamefont
  {Huerta}},\ }\href@noop {} {\bibfield  {journal} {\bibinfo  {journal}
  {Journal of Physics A: Mathematical and Theoretical}\ }\textbf {\bibinfo
  {volume} {42}},\ \bibinfo {pages} {504007} (\bibinfo {year}
  {2009})}\BibitemShut {NoStop}%
\bibitem [{\citenamefont {Singh}\ \emph {et~al.}(2012)\citenamefont {Singh},
  \citenamefont {Melko},\ and\ \citenamefont {Oitmaa}}]{Singh:2012t}%
  \BibitemOpen
  \bibfield  {author} {\bibinfo {author} {\bibfnamefont {R.~R.~P.}\
  \bibnamefont {Singh}}, \bibinfo {author} {\bibfnamefont {R.~G.}\ \bibnamefont
  {Melko}}, \ and\ \bibinfo {author} {\bibfnamefont {J.}~\bibnamefont
  {Oitmaa}},\ }\href@noop {} {\bibfield  {journal} {\bibinfo  {journal} {Phys.
  Rev. B}\ }\textbf {\bibinfo {volume} {86}},\ \bibinfo {pages} {075106}
  (\bibinfo {year} {2012})}\BibitemShut {NoStop}%
\bibitem [{\citenamefont {Kallin}\ \emph {et~al.}(2013)\citenamefont {Kallin},
  \citenamefont {Hyatt}, \citenamefont {Singh},\ and\ \citenamefont
  {Melko}}]{Kallin:2013}%
  \BibitemOpen
  \bibfield  {author} {\bibinfo {author} {\bibfnamefont {A.~B.}\ \bibnamefont
  {Kallin}}, \bibinfo {author} {\bibfnamefont {K.}~\bibnamefont {Hyatt}},
  \bibinfo {author} {\bibfnamefont {R.~R.~P.}\ \bibnamefont {Singh}}, \ and\
  \bibinfo {author} {\bibfnamefont {R.~G.}\ \bibnamefont {Melko}},\ }\href@noop
  {} {\bibfield  {journal} {\bibinfo  {journal} {Phys. Rev. Lett.}\ }\textbf
  {\bibinfo {volume} {110}},\ \bibinfo {pages} {135702} (\bibinfo {year}
  {2013})}\BibitemShut {NoStop}%
\bibitem [{\citenamefont {Kallin}\ \emph {et~al.}(2014)\citenamefont {Kallin},
  \citenamefont {Stoudenmire}, \citenamefont {Fendley}, \citenamefont {Singh},\
  and\ \citenamefont {Melko}}]{Kallin:2014}%
  \BibitemOpen
  \bibfield  {author} {\bibinfo {author} {\bibfnamefont {A.~B.}\ \bibnamefont
  {Kallin}}, \bibinfo {author} {\bibfnamefont {E.~M.}\ \bibnamefont
  {Stoudenmire}}, \bibinfo {author} {\bibfnamefont {P.}~\bibnamefont
  {Fendley}}, \bibinfo {author} {\bibfnamefont {R.~R.~P.}\ \bibnamefont
  {Singh}}, \ and\ \bibinfo {author} {\bibfnamefont {R.~G.}\ \bibnamefont
  {Melko}},\ }\href@noop {} {\bibfield  {journal} {\bibinfo  {journal} {Journal
  of Statistical Mechanics: Theory and Experiment}\ }\textbf {\bibinfo {volume}
  {2014}},\ \bibinfo {pages} {P06009} (\bibinfo {year} {2014})}\BibitemShut
  {NoStop}%
\bibitem [{\citenamefont {Helmes}\ and\ \citenamefont
  {Wessel}(2014)}]{Helmes:2014}%
  \BibitemOpen
  \bibfield  {author} {\bibinfo {author} {\bibfnamefont {J.}~\bibnamefont
  {Helmes}}\ and\ \bibinfo {author} {\bibfnamefont {S.}~\bibnamefont
  {Wessel}},\ }\href@noop {} {\bibfield  {journal} {\bibinfo  {journal} {Phys.
  Rev. B}\ }\textbf {\bibinfo {volume} {89}},\ \bibinfo {pages} {245120}
  (\bibinfo {year} {2014})}\BibitemShut {NoStop}%
\bibitem [{\citenamefont {Devakul}\ and\ \citenamefont
  {Singh}(2014{\natexlab{a}})}]{Devakul:2014}%
  \BibitemOpen
  \bibfield  {author} {\bibinfo {author} {\bibfnamefont {T.}~\bibnamefont
  {Devakul}}\ and\ \bibinfo {author} {\bibfnamefont {R.~R.~P.}\ \bibnamefont
  {Singh}},\ }\href@noop {} {\bibfield  {journal} {\bibinfo  {journal} {Phys.
  Rev. B}\ }\textbf {\bibinfo {volume} {90}},\ \bibinfo {pages} {064424}
  (\bibinfo {year} {2014}{\natexlab{a}})}\BibitemShut {NoStop}%
\bibitem [{\citenamefont {Devakul}\ and\ \citenamefont
  {Singh}(2014{\natexlab{b}})}]{Devakul:2014e}%
  \BibitemOpen
  \bibfield  {author} {\bibinfo {author} {\bibfnamefont {T.}~\bibnamefont
  {Devakul}}\ and\ \bibinfo {author} {\bibfnamefont {R.~R.~P.}\ \bibnamefont
  {Singh}},\ }\href@noop {} {\bibfield  {journal} {\bibinfo  {journal} {Phys.
  Rev. B}\ }\textbf {\bibinfo {volume} {90}},\ \bibinfo {pages} {054415}
  (\bibinfo {year} {2014}{\natexlab{b}})}\BibitemShut {NoStop}%
\bibitem [{\citenamefont {Inglis}\ and\ \citenamefont
  {Melko}(2013)}]{Inglis_2013}%
  \BibitemOpen
  \bibfield  {author} {\bibinfo {author} {\bibfnamefont {S.}~\bibnamefont
  {Inglis}}\ and\ \bibinfo {author} {\bibfnamefont {R.~G.}\ \bibnamefont
  {Melko}},\ }\href@noop {} {\bibfield  {journal} {\bibinfo  {journal} {New
  Journal of Physics}\ }\textbf {\bibinfo {volume} {15}},\ \bibinfo {pages}
  {073048} (\bibinfo {year} {2013})}\BibitemShut {NoStop}%
\bibitem [{\citenamefont {Zhang}\ \emph {et~al.}(2013)\citenamefont {Zhang},
  \citenamefont {Wierschem}, \citenamefont {Yap}, \citenamefont {Kato},
  \citenamefont {Batista},\ and\ \citenamefont {Sengupta}}]{Zhang:2013p}%
  \BibitemOpen
  \bibfield  {author} {\bibinfo {author} {\bibfnamefont {Z.}~\bibnamefont
  {Zhang}}, \bibinfo {author} {\bibfnamefont {K.}~\bibnamefont {Wierschem}},
  \bibinfo {author} {\bibfnamefont {I.}~\bibnamefont {Yap}}, \bibinfo {author}
  {\bibfnamefont {Y.}~\bibnamefont {Kato}}, \bibinfo {author} {\bibfnamefont
  {C.~D.}\ \bibnamefont {Batista}}, \ and\ \bibinfo {author} {\bibfnamefont
  {P.}~\bibnamefont {Sengupta}},\ }\href@noop {} {\bibfield  {journal}
  {\bibinfo  {journal} {Phys. Rev. B}\ }\textbf {\bibinfo {volume} {87}},\
  \bibinfo {pages} {174405} (\bibinfo {year} {2013})}\BibitemShut {NoStop}%
\bibitem [{\citenamefont {Sandvik}(1999)}]{Sandvik:1999}%
  \BibitemOpen
  \bibfield  {author} {\bibinfo {author} {\bibfnamefont {A.~W.}\ \bibnamefont
  {Sandvik}},\ }\href@noop {} {\bibfield  {journal} {\bibinfo  {journal} {Phys.
  Rev. B}\ }\textbf {\bibinfo {volume} {59}},\ \bibinfo {pages} {R14157}
  (\bibinfo {year} {1999})}\BibitemShut {NoStop}%
\bibitem [{\citenamefont {Pollock}\ and\ \citenamefont
  {Ceperley}(1987)}]{Pollock_Ceperley}%
  \BibitemOpen
  \bibfield  {author} {\bibinfo {author} {\bibfnamefont {E.~L.}\ \bibnamefont
  {Pollock}}\ and\ \bibinfo {author} {\bibfnamefont {D.~M.}\ \bibnamefont
  {Ceperley}},\ }\href {\doibase 10.1103/PhysRevB.36.8343} {\bibfield
  {journal} {\bibinfo  {journal} {Phys. Rev. B}\ }\textbf {\bibinfo {volume}
  {36}},\ \bibinfo {pages} {8343} (\bibinfo {year} {1987})}\BibitemShut
  {NoStop}%
\bibitem [{\citenamefont {Rigol}\ \emph {et~al.}(2006)\citenamefont {Rigol},
  \citenamefont {Bryant},\ and\ \citenamefont {Singh}}]{Rigol:2006}%
  \BibitemOpen
  \bibfield  {author} {\bibinfo {author} {\bibfnamefont {M.}~\bibnamefont
  {Rigol}}, \bibinfo {author} {\bibfnamefont {T.}~\bibnamefont {Bryant}}, \
  and\ \bibinfo {author} {\bibfnamefont {R.~R.~P.}\ \bibnamefont {Singh}},\
  }\href@noop {} {\bibfield  {journal} {\bibinfo  {journal} {Phys. Rev. Lett.}\
  }\textbf {\bibinfo {volume} {97}},\ \bibinfo {pages} {187202} (\bibinfo
  {year} {2006})}\BibitemShut {NoStop}%
\bibitem [{\citenamefont {Rigol}\ \emph
  {et~al.}(2007{\natexlab{a}})\citenamefont {Rigol}, \citenamefont {Bryant},\
  and\ \citenamefont {Singh}}]{Rigol:2007_1}%
  \BibitemOpen
  \bibfield  {author} {\bibinfo {author} {\bibfnamefont {M.}~\bibnamefont
  {Rigol}}, \bibinfo {author} {\bibfnamefont {T.}~\bibnamefont {Bryant}}, \
  and\ \bibinfo {author} {\bibfnamefont {R.~R.~P.}\ \bibnamefont {Singh}},\
  }\href@noop {} {\bibfield  {journal} {\bibinfo  {journal} {Phys. Rev. E}\
  }\textbf {\bibinfo {volume} {75}},\ \bibinfo {pages} {061118} (\bibinfo
  {year} {2007}{\natexlab{a}})}\BibitemShut {NoStop}%
\bibitem [{\citenamefont {Rigol}\ \emph
  {et~al.}(2007{\natexlab{b}})\citenamefont {Rigol}, \citenamefont {Bryant},\
  and\ \citenamefont {Singh}}]{Rigol:2007_2}%
  \BibitemOpen
  \bibfield  {author} {\bibinfo {author} {\bibfnamefont {M.}~\bibnamefont
  {Rigol}}, \bibinfo {author} {\bibfnamefont {T.}~\bibnamefont {Bryant}}, \
  and\ \bibinfo {author} {\bibfnamefont {R.~R.~P.}\ \bibnamefont {Singh}},\
  }\href@noop {} {\bibfield  {journal} {\bibinfo  {journal} {Phys. Rev. E}\
  }\textbf {\bibinfo {volume} {75}},\ \bibinfo {pages} {061119} (\bibinfo
  {year} {2007}{\natexlab{b}})}\BibitemShut {NoStop}%
\bibitem [{\citenamefont {Tang}\ \emph {et~al.}(2013)\citenamefont {Tang},
  \citenamefont {Khatami},\ and\ \citenamefont {Rigol}}]{Tang:2013}%
  \BibitemOpen
  \bibfield  {author} {\bibinfo {author} {\bibfnamefont {B.}~\bibnamefont
  {Tang}}, \bibinfo {author} {\bibfnamefont {E.}~\bibnamefont {Khatami}}, \
  and\ \bibinfo {author} {\bibfnamefont {M.}~\bibnamefont {Rigol}},\
  }\href@noop {} {\bibfield  {journal} {\bibinfo  {journal} {Computer Physics
  Communications}\ }\textbf {\bibinfo {volume} {184}},\ \bibinfo {pages} {557 }
  (\bibinfo {year} {2013})}\BibitemShut {NoStop}%
\bibitem [{\citenamefont {Schollw{\"o}ck}(2005)}]{Schollwoeck:2005}%
  \BibitemOpen
  \bibfield  {author} {\bibinfo {author} {\bibfnamefont {U.}~\bibnamefont
  {Schollw{\"o}ck}},\ }\href@noop {} {\bibfield  {journal} {\bibinfo  {journal}
  {Rev. Mod. Phys.}\ }\textbf {\bibinfo {volume} {77}},\ \bibinfo {pages} {259}
  (\bibinfo {year} {2005})}\BibitemShut {NoStop}%
\bibitem [{ITe()}]{ITensor}%
  \BibitemOpen
  \href@noop {} {}\bibinfo {note} {\mbox{C}alculations were performing using
  the ITensor Library: http://itensor.org/}\BibitemShut {NoStop}%
\bibitem [{\citenamefont {Stoudenmire}\ and\ \citenamefont
  {White}(2012)}]{Stoudenmire}%
  \BibitemOpen
  \bibfield  {author} {\bibinfo {author} {\bibfnamefont {E.~M.}\ \bibnamefont
  {Stoudenmire}}\ and\ \bibinfo {author} {\bibfnamefont {S.~R.}\ \bibnamefont
  {White}},\ }\href@noop {} {\bibfield  {journal} {\bibinfo  {journal} {Annual
  Review of Condensed Matter Physics}\ }\textbf {\bibinfo {volume} {3}},\
  \bibinfo {pages} {111} (\bibinfo {year} {2012})}\BibitemShut {NoStop}%
\bibitem [{\citenamefont {Kallin}(2014)}]{Kallin_thesis}%
  \BibitemOpen
  \bibfield  {author} {\bibinfo {author} {\bibfnamefont {A.~B.}\ \bibnamefont
  {Kallin}},\ }\emph {\bibinfo {title} {Computational Methods for the
  Measurement of Entanglement in Condensed Matter Systems}},\ \href
  {https://uwspace.uwaterloo.ca/handle/10012/8539} {Ph.D. thesis},\ \bibinfo
  {school} {University of Waterloo} (\bibinfo {year} {2014})\BibitemShut
  {NoStop}%
\bibitem [{\citenamefont {Senthil}\ \emph {et~al.}(2004)\citenamefont
  {Senthil}, \citenamefont {Vishwanath}, \citenamefont {Balents}, \citenamefont
  {Sachdev},\ and\ \citenamefont {Fisher}}]{Senthil:2004}%
  \BibitemOpen
  \bibfield  {author} {\bibinfo {author} {\bibfnamefont {T.}~\bibnamefont
  {Senthil}}, \bibinfo {author} {\bibfnamefont {A.}~\bibnamefont {Vishwanath}},
  \bibinfo {author} {\bibfnamefont {L.}~\bibnamefont {Balents}}, \bibinfo
  {author} {\bibfnamefont {S.}~\bibnamefont {Sachdev}}, \ and\ \bibinfo
  {author} {\bibfnamefont {M.~P.~A.}\ \bibnamefont {Fisher}},\ }\href@noop {}
  {\bibfield  {journal} {\bibinfo  {journal} {Science}\ }\textbf {\bibinfo
  {volume} {303}},\ \bibinfo {pages} {1490} (\bibinfo {year}
  {2004})}\BibitemShut {NoStop}%
\bibitem [{\citenamefont {Sandvik}(2010)}]{Sandvik:2010d}%
  \BibitemOpen
  \bibfield  {author} {\bibinfo {author} {\bibfnamefont {A.~W.}\ \bibnamefont
  {Sandvik}},\ }\href@noop {} {\bibfield  {journal} {\bibinfo  {journal} {Phys.
  Rev. Lett.}\ }\textbf {\bibinfo {volume} {104}},\ \bibinfo {pages} {177201}
  (\bibinfo {year} {2010})}\BibitemShut {NoStop}%
\bibitem [{\citenamefont {Block}\ \emph {et~al.}(2013)\citenamefont {Block},
  \citenamefont {Melko},\ and\ \citenamefont {Kaul}}]{Block_2013}%
  \BibitemOpen
  \bibfield  {author} {\bibinfo {author} {\bibfnamefont {M.~S.}\ \bibnamefont
  {Block}}, \bibinfo {author} {\bibfnamefont {R.~G.}\ \bibnamefont {Melko}}, \
  and\ \bibinfo {author} {\bibfnamefont {R.~K.}\ \bibnamefont {Kaul}},\ }\href
  {http://link.aps.org/doi/10.1103/PhysRevLett.111.137202} {\bibfield
  {journal} {\bibinfo  {journal} {Phys. Rev. Lett.}\ }\textbf {\bibinfo
  {volume} {111}},\ \bibinfo {pages} {137202} (\bibinfo {year}
  {2013})}\BibitemShut {NoStop}%
\end{thebibliography}%

\end{document}